\documentclass[journal]{IEEEtran}
\usepackage{cite}
\usepackage{array}
\usepackage{latexsym,amsmath,amssymb,graphicx}
\begin{document}
\title{Design of Nested LDGM-LDPC Codes for Compress-and-Forward in Relay Channel}
\author{Bingbing Zheng, Lingge~Jiang, Chen~He and Qingchuan Wang\\
Department of Electronic Engineering\\
Shanghai Jiao Tong University\\
Email: doubleice@sjtu.edu.cn
\thanks{This paper was supported by National Natural Science Foundation of
China Grants No.60872017 and State 863 Project of China Grants
No.2009AA011505}}

\markboth{}{Shell \MakeLowercase{\textit{et al.}}: Bare Demo of
IEEEtran.cls for Journals}
 \maketitle
\begin{abstract}
A three terminal relay system with binary erasure channel (BEC)
was considered, in which a source forwarded information to a
destination with a relay's "assistance". The nested LDGM
(Low-density generator-matrix) -LDPC (low-density parity-check)
was designed to realize Compress-and-forward (CF) at the relay.
LDGM coding compressed the received signals losslessly and LDPC
realized the binning for Slepian-Wolf coding. Firstly a practical
coding scheme was proposed to achieve the cut-set bound on the
capacity of the system, employing LDPC and Nested LDGM-LDPC codes
at the source and relay respectively. Then, the degree
distribution of LDGM and LDPC codes was optimized with a given
rate bound, which ensured that the iterative belief propagation
(BP) decoding algorithm at the destination was convergent.
Finally, simulations results show that the performance achieved
based on nested codes is very close to Slepian-Wolf theoretical
limit.
\end{abstract}

\section{Introduction}
\PARstart{C}{ooperative} communication \cite{1} has recently
attracted much attention due to it can achieve a larger rate
region, compared with traditional networks. Many cooperative
protocols have been proposed in the literature. These protocols
are usually classified into three categories: decode-and-forward
(DF) protocol, where the relay decodes and re-encodes the signals
transmitted by the source; amplify-and-forward (AF) protocol, in
which a relay simply amplifies its received signals;
compress-and-forward (CF) protocol, where the relay compresses the
signals from the source, and forwards these compressed soft
information to the destination. For DF protocol, it has been
widely researched based on LDPC codes \cite{2}, and suffers a loss
of performance when the relay can't be guaranteed to recover the
source information. The relay always can assist the source to
convey information with amplified soft information when AF
protocol is employed, but the protocol is suboptimal. CF protocol,
jointing source-channel coding, is a form of Wyner-Ziv (WZ) coding
\cite{3} in case of lossy compression and Slepian-Wolf coding
\cite{4} in the lossless case. It takes advantages of the
statistical dependence of the relay's and destination's channel
output, and achieves higher rate than DF and AF.
\\
\indent So far, most of the researches concerning CF remain at the
theoretical level and the realization of CF is tackled in only a
few papers, \cite{5}\cite{6}\cite{7}. These papers achieve
Slepian-Wolf compression at the relay by taking the syndrome of an
LDPC code \cite{5}, or by using an Irregular Repeat Accumulator
(IRA) code \cite{6} that combines Slepian-Wolf coding with channel
coding on the relay-to-destination channel. However these
approaches are suitable only when the source-relay channel output
is binary and the relay does lossless (rather than lossy)
compression. Otherwise, e.g. the Gaussian relay channel is
considered in \cite{7}, significant capacity losses will result.
\\
\indent It is known that LDPC codes are good channel codes and
recent work has also shown that LDGM codes are good source codes.
Some near-ideal encoding/decoding algorithms with LDGM codes have
been proposed \cite{8}\cite{9}, furthermore nested LDGM-LDPC codes
often are used to guarantee both channel coding and source coding
performance. In \cite{10}, such nested codes are used to approach
capacity in dirty paper coding, where good channel coding ensures
low error probability and good source coding guarantees good
shaping of the transmitted signal.
\\
\indent To focus on the effective compression of the received data
at the relay, we consider a three terminal cooperative system
where the source-destination and source-relay links are both
binary erasure channel (BEC), and the relay-destination link is
orthogonal to them. As the received signal by the relay through
the BEC is 3-ary, the aforementioned syndrome methods are
insufficient, making the proposed nested codes necessary.
\\
\indent This paper is organized as follows. Section II gives a
brief introduction to the system model involving a relay channel.
Section III describes the compression and decoding algorithms
based on nested LDGM-LDPC codes. Section IV presents the degree
distribution optimization method necessary for good performance.
Experiment results are given in Section V to verify the
effectiveness of the proposed algorithms and optimization methods.
Finally, Section VI concludes the paper.
\section{system model}
The full-duplex single relay system is shown in Fig.1. It
comprises a source S, a destination D and a relay node R. The S-R
and S-D links, both with erasure probability $\varepsilon$, form a
binary erasure broadcast channel. The R-D link, with a capacity
denoted by $C_{rd}$, is orthogonal to the S-D and S-R links.
Denoting an erased bit by $E$, the relay system is described by
four random variables $x_s$, $x_r$, $y_r$, $y_d$ and one
conditional probability distribution $p(y_r,y_d|x_s)$,which is
shown in Table I.
\begin{figure}
\begin{center}
\includegraphics[width=2in]{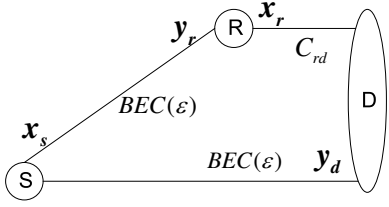}
\caption{The single relay system} \label{fig.1}
\end{center}
\end{figure}
\begin{table}[!hbp]
\caption{Conditional probability distribution of
$p(y_r,y_d|x_s=0)$, When $p(y_r,y_d|x_s=1)$, the variables value:
0 and 1 exchange} \label{table_1} \centering
\begin{tabular}{cccc}
\hline $y_d,y_r$ & 0& 1 & $E$ \\

 \hline
0& $(1-\varepsilon)^2$ &0 &$(1-\varepsilon)\varepsilon$ \\
1&0&0&0\\
$E$ &$(1-\varepsilon)\varepsilon$ &0&$\varepsilon^2$\\
 \hline
\end{tabular}
\end{table}

\indent In every block a message $w$ which is random variable
uniformly distributed on $[1,2^{nR})$ is encoded into sequence
$\boldsymbol{x_s}=(x_{s1},x_{s2},\cdots,x_{sn})$ at the source S,
and transmitted through BEC.
$\boldsymbol{y_r}=(y_{r1},y_{r2},\cdots,y_{rn})$,
$\boldsymbol{y_d}=(y_{d1},y_{d2},\cdots,y_{dn})$ is received by R
and D respectively. At R $\boldsymbol{y_r}$ is decoded or
loselessly compressed, and then re-encoded into
$\boldsymbol{x_r}=(x_{r1},x_{r2},\cdots,x_{rn'})$ to implement DF
or CF respectively. When $C_{rd}$ is sufficiently large,
$\boldsymbol{y_r}$ could be decoded with the compressed signals
$\boldsymbol{x_r}$ and the side information $\boldsymbol{y_d}$.
$w$ could be recovered at D by jointly decoding $\boldsymbol{y_r}$
and $\boldsymbol{y_d}$. Then the achievable rate \cite{11} of the
system is given by\\
1) If $\boldsymbol{y_r}$ is decoded, then
\begin{equation}
R \leq I(x_s;y_r|x_r)=1-\varepsilon
\end{equation}
2) If $\boldsymbol{y_r}$ is compressed and encoded into
$\boldsymbol{x_r}$, with large $C_{rd}$ the compression can be
lossless, in which case
\begin{equation}
R \leq I(x_s;y_ry_d)=1-\varepsilon^2
\end{equation}
It can be seen from (1) that, when $C_{rd}$ is large, the S-R link
become the bottleneck of DF mode. However CF can achieve a higher
rate in (2), which is actually the cut-set bound \cite{11} on the
capacity of the relay channel. On the other hand, in CF mode the
relay node is unable to decode $\boldsymbol{y_r}$ , so the
$C_{rd}$ must be large enough to transmit all the information
about the erased positions.\\
\indent In this paper, we focus on lossless CF, and since it
already achieves capacity, the remaining task is to minimize
$C_{rd}$ the necessary. As $\boldsymbol{y_d}$ is available at D
and is correlated with $\boldsymbol{y_r}$, it can be used as side
information to reduce $\boldsymbol{y_r}$'s encoding rate from
$H(y_r)$ into $H(y_r|y_d)$ through Slepian-Wolf coding. In the
next section, nested LDGM-LDPC codes will be designed to realize
the compression and binning necessary for Slepian-Wolf coding.
\section{The nested LDGM-LDPC code for compression}
Considering that LDPC as channel code achieves
capacity-approaching performance with low-complexity iterative
decoding manner, we encode the source messages with an LDPC code,
denoted by factor graph \cite{12} $C_1$, which is shown in Fig.2.
The circles denote the variable nodes q, representing $n$ bits
codeword of LDPC codes, and the black squares denote the check
(function) nodes s, representing  $k$  parity check equations. The
number of edges connected to one node is denoted as the degree of
the node. The rate $R_0=\frac{n-k}{n}$  could approach cut-set
bound in (2) by optimizing the degree distribution of $C_1$, which
will is discussed in the next section.
\begin{figure}
\begin{center}
\includegraphics[width=0.8in]{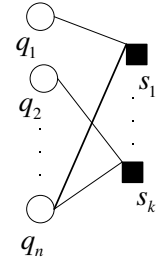}
\caption{The LDPC code $C_1$} \label{fig.2}
\end{center}
\end{figure}
\indent With $C_1$ the message $w$ at S is encoded into a binary
sequence $\boldsymbol{x_s}\in\{0,1\}^n$ and transmitted over BEC.
Then the received sequence $\boldsymbol{y_r}$ by R is relayed to D
via CF. We design a nested LDGM-LDPC construction $C_2$, to deal
with the compression problem at R, whose factor graph is shown in
Fig.3. The circles are also the variable nodes and black squares
are the factor nodes. As each is regarded as a ternary symbol, we
firstly map $\boldsymbol{y_r}$ into a binary sequence
$\boldsymbol{c}$ by
\begin{eqnarray}
\hat{y}_{ri}=\varphi(c_{2i-1}\oplus v_{2i-1} \oplus
\zeta_{2i-1},c_{2i}\oplus v_{2i} \oplus \zeta_{2i}), \nonumber \\
\varphi(00)=0,\varphi(01)=1,\varphi(1*)=E,i=1,2,\cdots,n,
\end{eqnarray}
Where $*$ denotes "don't care" positions that can be encoded into
either 0 or 1. $\boldsymbol{v}$ is a pseudo-random dither sequence
which is added to assure uniform distribution.
$\boldsymbol{\zeta}$ is normally an all-zero sequence, but in
practice erroneous decimation of the b-nodes will inevitably occur
and cause contradictions, which must be corrected by flipping the
bits in corresponding to the c-nodes with contradictions. Here the
factor nodes connected to  $\boldsymbol{y_r}$ and $\boldsymbol{c}$
represent the mapping in (3). Then with LDGM part of $C_2$ , the
binary sequence containing $*$ is quantized into a shorter one
$\boldsymbol{b}$ by
\begin{eqnarray}
c=bG,b\in\{0,1\}^m,m=nR_b
\end{eqnarray}
where $G$ is the generation matrix and $R_b$ is optimized to be
slightly larger than $2-\varepsilon$ , so that LDGM coding with
$R_b$ is lossless. After that, we use the LDPC part of $G_2$ to
compress $\boldsymbol{b}$ into $\boldsymbol{p}$, with
\begin{eqnarray}
p=bH^T,p\in\{0,1\}^t,t=nR_p
\end{eqnarray}
where $H$ is the sparse parity check matrix and $R_p$ also
optimized is slightly larger than $H(y_r|y_d)$. $H(y_r|y_d)$ is
the Slepian-Wolf theoretical limit.\\
\begin{figure}
\begin{center}
\includegraphics[width=2.5in]{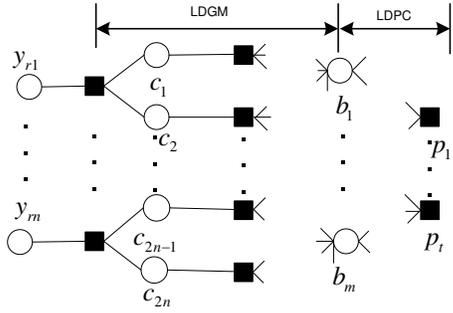}
\caption{The nested LDGM-LDPC codes $C_2$} \label{fig.3}
\end{center}
\end{figure}
\indent Until now the nested LDGM-LDPC code has been used to
compress $\boldsymbol{y_r}$ into $\boldsymbol{p}$ (as well as the
flipped positions $\boldsymbol{\zeta}$) at the rate approximately
equal to $H(y_r|y_d)$. Assuming that $C_{rd}$ is sufficient for
transmitting these information, D can decode $\boldsymbol{y_r}$
from the side information $\boldsymbol{y_r}$ and the compressed
information $\boldsymbol{p}$ losslessly. The iterative belief
propagation \cite{13} decoding process is also executed from
Fig.3. Finally, combining $\boldsymbol{y_r}$ with
$\boldsymbol{y_d}$, D recovers the message $w$, again by the BP
algorithm.
\section{Code Optimization $C_1$ and $C_2$}
When LDPC, LDGM and nested codes are designed, it is always
critical to optimize the degree distributions so that BP converges
well, which can be visualized on the EXIT chart \cite{14} as a
gap. So we optimize the degree distribution using the EXIT and EBP
curve (Extended BP curve which is another form of EXIT
chart)\cite{15} considering both the encoding process and decoding
process also from the source node, relay node and destination
node.\\
\indent  The design of $C_1$ involves just LDPC optimization over
a BEC with erasure probability $\varepsilon^2$ , which means that
the prior information at q-nodes acquired from channel output
becomes $I_{q,pri}=1-\varepsilon^2$. S-regular, q-irregular LDPC
code \cite{13} is designed to achieve good performance. Let $d_s$
be the left-degree of all s-nodes, and denote $v_{qd}$ as the
fraction of edges connected to q-nodes with the right-degree $d$.
$I_{qs}$, $I_{sq}$ represent the average mutual information (MI)
in every q-to-s, s-to-q message at a certain iteration
respectively. Then the optimization the degree distribution
$v_{qd}$ of q-nodes is a linear programming problem, which is
\begin{align}
&max R_0=1-\frac{1/d_s}{\sum_{d}v_{qd}/d}\nonumber\\
&s.t.\sum_{d}v_{qd}=1,I_{qs}>I_{qs}^{-}+\vartriangle_{qs},I_{sq}\in[0,1]
\end{align}
in which
\begin{eqnarray}
I_{qs}=1-(1-I_{q,pri})\sum_{d}v_{qd}(1-I_{sq})^{d-1}
\end{eqnarray}
\begin{eqnarray}
I_{qs}^{-}=(I_{sq})^{\frac{1}{d_s-1}}
\end{eqnarray}
where superscript "-" refers to last iteration. EXIT curves at
q-nodes and s-nodes called q-curve and s-curve are given by (7)
and (8). $\vartriangle_{qs}$ is added to ensure that there are
some gap between the matched EXIT curves, so that BP algorithm
won't get stuck.\\
\indent For the nested LDGM-LDPC code $C_2$, the LDGM part is
essentially dictated by good encoder-side performance at the relay
node, so we optimize it first. As Ternary symbol $y_{ri}$ is
encoded into two bits $c_{2i-1}$,$c_{2i}$ in c-nodes, to simplify
analysis, we assume all b-nodes have the left-degree $d_b$ and the
two bits $c_{2i-1}$,$c_{2i}$ connected to the same $y_{ri}$-node
have the same right-degree, called the c-degree of the $y_r$-node.
Now we only have to optimize the c-degree distribution of the
$y_r$-nodes, represented by $v_{cd}$, the fraction of edges
connected to $y_r$-nodes with c-degree $d$ from the edge
perspective. Besides, the optimization is needed with the
constraint of the monotonic condition \cite{9}, which makes sure
that encoding can proceed with a vanishing fraction of
contradictions and thus flipped bits. Thus the optimization
problem is summarized as
\begin{align}
&\min_{v_{cd}}R_{b}=\frac{2}{d_b\sum_{d}v_{cd}/d}\nonumber\\
&s.t.\sum_{d}v_{cd}=1,I_{bc,pri}|_{I_{bc}=0}\geq 0,
\frac{dI_{bc,pri}}{dI_{bc}}\geq 0, I_{bc}\in[0,1]
\end{align}
in which
\begin{eqnarray}
&I_{bc,pri}=1-(1-I_{bc})/(1-I_{cb})^{d_b-1}
\end{eqnarray}
\begin{eqnarray}
&I_{cb}=I_{yc}\sum_dv_{cd}(I_{bc})^{d-1}
\end{eqnarray}
Where $I_{bc}$, $I_{cb}$ denotes the average MI in every b-to-c,
c-to-b message at a certain iteration respectively, and
$I_{yc}=1-0.5\varepsilon$ represents the priors average MI of the
$y_r$-nodes with c-degree, which is acquired from the mapping in
(3). $I_{bc,pri}$ denotes the priors average MI of the b-nodes at
fixed points(i.e. $I_{bc,pri}$ making the average MI
$I_{bc}=I_{bc}^{-}$), and it should be 0 when $I_{bc}=0$ and
increase monotonically as $I_{bc}$ increase from 0 to 1.\\
\indent With the degree distribution of LDGM part fixed, we
optimize the degree distribution of the LDPC part of $C_2$ to
achieve decoder-side performance. During the iterative decoding of
$\boldsymbol{y_r}$ at the destination, the average MI of the
message from $y_r$- to c-nodes, denoted as $I_{yc,d}$ for those
with c-degree $d$, varies between $I_{c0}=I(c;y_d)$ and
$I_{c1}=I(c_{2i-1};y_{di}|c_{2i})+I(c_{2i};y_{di}|c_{2i-1})$,
according to the incoming messages from c- and -nodes, that is
\begin{eqnarray}
I_{yc,d}=I_{c0}(1-I_{bc}^d)+I_{c1}I_{bc}^d
\end{eqnarray}
Making $I_{cb}$ in (11) become
\begin{eqnarray}
I_{cb}=\sum_dv_{cd}I_{yc,d}(I_{bc})^{d-1}
\end{eqnarray}
Let $I_{bc,ext}$ denote the extrinsic MI of b-node at fixed
points, derived only from $I_{cb}$, which is
\begin{eqnarray}
I_{bc,ext}=1-(1-I_{cb})^{d_b}
\end{eqnarray}
Thus the decoder-side EBP curve of the LDGM part formed by
$I_{bc,pri}$ vs. $I_{bc,ext}$ is derived from (10) and (14). The
LDPC part is designed to make the EBP curve of LDPC part
$I_{bp,pri}$ vs. $I_{bp,ext}$ match that of LDGM part, so that
$\boldsymbol{y_r}$ can be decoded. In other words, suppose the EBP
curve is plotted with $I_{bp,ext}$ in the horizontal axis and
$I_{bc,ext}$ in the vertical axis, then EBP curve of the LDPC part
should lie below that of the LDGM part, with a small gap between
them. The gap assures that iterative decoding does not get stuck.
Thus let $v_{pd}$ and $v_{bd}$ denotes the fraction of edges
connected to p-node and b-node with the left-degree and
right-degree $d$ respectively. The degree distributions are
optimized to achieve the minimal rate, which is
\begin{align}
&\min_{v_{pd},v_{bd}}R_{p}=R_b-R_{bp}=R_b-(1-\frac{\sum_{d}v_{pd}/d}{\sum_{d}v_{bd}/d})\nonumber\\
&s.t.\sum_{d}v_{pd}=1,\sum_{d}v_{bd}=1,I_{pb}^{+}>I_{pb}+\vartriangle_{pb},I_{pb}\in[0,1]\nonumber\\
&I_{bp,ext}=1-\sum_{d}v_{bd}(1-I_{pb})^d \nonumber\\
&I_{bc,pri}=I_{bp,ext},I_{bp,pri}=I_{bc,ext}\nonumber\\
&I_{bp}=1-(1-I_{bp,pri})\sum_{d}v_{bd}(1-I_{pb})^{d-1}\nonumber\\
&I_{pb}^{+}=\sum_{d}v_{pd}(I_{bp})^{d-1}
\end{align}
Where $I_{bc,ext}$ is derived from the EBP curve of LDGM part at
the decoder-side with the corresponding $I_{bc,pri}$ known.
$I_{bp}$, $I_{pb}$ denotes the average MI in every b-to-p, p-to-b
message at a certain iteration respectively, and superscript "+"
refers to the next iteration. $\vartriangle_{pb}$ is designed to
keep the gap in the EBP curves of the LDGM and LDPC part to make
the BP converge.
\section{Numerical Results}
In this section we evaluate the performance of our optimization
for degree distribution and LDGM-LDPC encoding and decoding
process. Let the erasure probabilities of the S-R and S-D links be
$\varepsilon=0.5$, the R-D link be ideal with $C_{rd}$ at least
1.25 bit/sym, and the block length $n=10^5$. Since the cut-set
bound in (2) is $I(x_s;y_dy_d)=0.75$ bit/sym, the code rate of
$C_1$ is $R_0\leq0.75$ bit/sym. The degree distribution of $C_1$
is optimized to achieve the maximal rate $R_0$. With
$2-\varepsilon=1.5$ bit/sym and $H(y_r|y_d)=1.25$ bit/sym, the
nested code $C_2$ has $R_b\geq1.5$ bit/sym, $R_p\geq1.25$
bit/sym.\\
\indent Optimizing the degree distribution of LDPC code by (6)
with $I_{q,pri}=0.75$, $d_s=16$, $R_0=0.742$ is acquired, and the
optimized degree distribution $v_{qd}$ is shown in Table II. With
$d_b=6$, $I_{yc}=0.75$, the degree distribution of LDGM part of
$C_2$ is optimized by (9),  and $R_b=1.5019$ is acquired. The
optimized degree distribution $v_{cd}$ of LDGM part at encoding
side is represented in Table III.
\begin{table}[!hbp]
\caption{The degree distribution of LDPC code $C_1$ at the
source} \label{table_2} \centering
\begin{tabular}{cccccccccc} \hline $d$ &
$v_{qd}$& $d$ & $v_{qd}$& $d$ & $v_{qd}$& $d$ &
$v_{qd}$& $d$ & $v_{qd}$ \\
\hline
2&0.2467&5&0.0154&8&0.0679&13&0.0027&21&0.0689\\
3&0.1768&5&0.0473&9&0.046 &17&0.0067&24&0.0835\\
4&0.0479&6&0.0712&10&0.018&19&0.0412&27&0.0598\\
 \hline
\end{tabular}
\end{table}

\begin{figure}
\begin{center}
\includegraphics[width=3in]{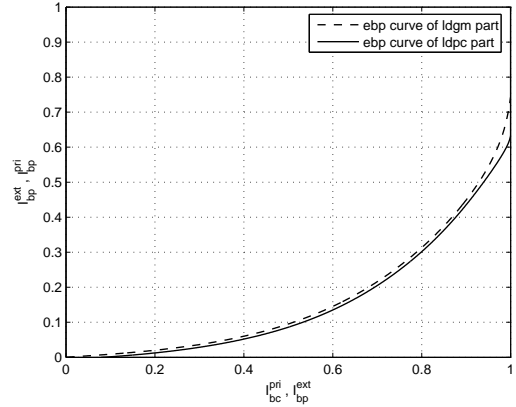}
\caption{The EBP curve of nested LDGM-LDPC codes in $C_2$ with gap
at destination node} \label{fig.4}
\end{center}

\end{figure}
\begin{table}[!hbp]
\caption{The degree distribution $v_{cd}$ of LDGM part in $C_2$}
\label{table_3} \centering
\begin{tabular}{cccccccccc} \hline $d$ &
$v_{cd}$& $d$ & $v_{cd}$& $d$ & $v_{cd}$& $d$ &
$v_{cd}$& $d$ & $v_{cd}$ \\
\hline
1&0.002 &6&0.026  &11&0.0081&21&0.0033&37&0.0016\\
2&0.5987&7&0.0161 &13&0.0074&24&0.0025&17&0.0013\\
3&0.1598&8&0.0126 &15&0.0056&27&0.0021&19&0.0012\\
4&0.0175&9&0.0099 &17&0.0047&30&0.0019&19&0.001\\
3&0.0408&10&0.0089&19&0.0043&33&0.0018& & \\
 \hline
\end{tabular}
\end{table}
\indent With $I_{c0}=0.0472$ and $I_{c1}=0.2028$, The EBP curve of
LDGM part at the decoder-side is shown in the dashed curve of
Fig.4. Then by (15), $R_p=1.2696$ is acquired and the optimized
degree distributions $v_{pd}$, $v_{bd}$ of LDPC part at the
decoder-side is shown in Table IV. The EBP curve of LDPC part is
shown in the solid curve of Fig.4 with some gap. It can be seen
that The EBP curve of LDPC part indeed lies below that of LDGM
part and both of them match well, which assure that the iterative
BP algorithm converge.\\
\begin{table}[!hbp]
\caption{The degree distribution $v_{bd}$,$v_{pd}$ of LDPC part in
$C_2$} \label{table_4} \centering
\begin{tabular}{cccccccccc}
\hline $d$ &$v_{bd}$& $d$ & $v_{bd}$& $d$ & $v_{bd}$& $d$ &
$v_{bd}$& $d$ & $v_{pd}$ \\
\hline
1&0.0039&4&0.0173 &7&0.1917&21&0.0947&2&0.6087\\
2&0.6505&7&0.0009 &13&0.0303&24&0.0108&17&0.3913\\
 \hline
\end{tabular}
\end{table}

\begin{table}[!hbp]
\caption{BER performance at destination: BEC channel; single-relay
system} \label{table_1} \centering
\begin{tabular}{ccc}
\hline Relay Protocol &Designed Rate & BER \\
\hline
CF&$R=0.472$&$1.357\times10^{-5}$\\
DF&$R=0.49$ &$3.265\times10^{-5}$\\
 \hline
\end{tabular}
\end{table}
\indent With optimized degree distribution of $C_1$ and $C_2$, the
simulation of encoding and decoding processes is executed. The
experiment result shows that the iteration count of BP decoding
the LDGM part in $C_2$ is about 200, and LDPC decoding in
destination is only about 150. However, when erasure probability
$\varepsilon$ decreases, which denotes the channel capacity
increases, the iteration count will decrease. With Monte Carlo
simulation, The BER performance at destination node for relay
system in BEC under CF and DF is shown in Table V. Simulation
shows that under CF the BER is about $10^5$ in most blocks, and
some blocks even recover source information correctly, which sees
that CF is much better than DF ($R\leq0.5$, by only optimizing the
degree distribution of $C_1$ to realize DF). Besides, $R_0=0.472$
is close to CF theoretical limit 0.75.\\
\indent Some remarks on the design of $C_1$ and $C_2$ are in
order. Firstly, when the degree distribution of LDPC in $C_1$ and
LDGM in $C_2$ is optimized, the degree of s-nodes and b-nodes
should be chosen carefully. Here reasonable choice of $d_s$ ranges
from 16 to 20 and $d_b=6$. Secondly, we should leave a uniform gap
between the EXIT curves of LDPC codes in $C_1$ and between EBP
curves of the LDGM and LDPC parts in $C_2$ to make iterative
decoding converge with a reasonable number of iterations rather
than getting stuck. Besides, in order not to cut down the designed
rate, we need assure the gap $\vartriangle_{qs}$ and
$\vartriangle_{pb}$ should not be larger than 0.01 in every
iteration. E.g. this gap of the EBP curves of the nested LDGM-LDPC
part in $C_2$ is designed with $\vartriangle_{pb}=0.004$, which is
shown in Fig.4. The decoding process of $\boldsymbol{y_r}$ will
not get stuck and with the flipped position known in the
destination, $\boldsymbol{y_r}$ could be decoded correctly.\\
\indent Thirdly, there will inevitably be some incorrect
decimation in the LDGM quantization process, which cause
contradictions that must be corrected by flipping some bits in
$\boldsymbol\zeta$. This $\boldsymbol\zeta$  must also be
transmitted to the destination using a fraction of $C_{rd}$, so
that it can perform decoding correctly. We found that the number
of flipped positions is about 600, and $R_p$ is 1.2696bit/sym, so
we require $C_{rd}=R_p+2*H_{2}(600/2*10^5)=1.8385$ bit/sym (here
$H_{2}(p)=p\log_{2}p+(1-p)\log_{2}(1-p)$). We could see that even
with the flipped positions transmitted the required channel
capacity $C_{rd}$ of R-D link can still be lower than $H(y_r)=1.5$
bit/sym. In our future work, we will study the lossy compression
of $\boldsymbol{y_r}$, so that we can get lower $C_{rd}$.\\
\indent It is also observed that the performance of the proposed
practical CF scheme improves as the block length increases at the
cost of larger memory consumption and coding delay.
\section{conclusion}
In this paper, a first practical CF scheme for a type of relay
system based on nested LDGM-LDPC has been proposed, and methods
for optimizing the degree distributions have been described.
Simulation results show that the nested LDGM-LDPC codes can
perform Slepian-Wolf compression of the relay's ternary received
signals when the relay system is BEC. The performance of our
scheme approaches the CF theoretical cut-set bound, while previous
schemes are either limited to binary signals or suboptimal. Our
work shows nested LDGM-LDPC codes for practical CF scheme is
sufficient.\\
\indent It is apparently straightforward to extend the proposed
scheme to realize lossy compression, which would offer better
performance achieved at a lower relay-destination channel capacity
. The design will be considered in our future work. We will also
try to optimize the gap between the BEP curves of the LDGM and
LDPC parts of the nested code, so that its decoding can converge
more quickly and reliably.

\end{document}